# Enhanced ferroelectric properties of low-annealed SrBi$_2$(Ta,Nb)$_2$O$_9$ thin films for NvFeRAM applications

O. M. Fesenko[1], A. D. Yaremkevich[1], T.V. Tsebrienko [1], O.P. Budnyk[1], Lei Wang[2], A. V. Semchenko[2], V. V. Sidski[2], A. N. Morozovska[1],

[1]*Institute of Physics, National Academy of Sciences of Ukraine, 46, pr. Nauky, 03028 Kyiv, Ukraine*

[2]*Nanjing University of Science and Technology, China, Jiangsu, Nanjing, Xuanwu, Xiaolingweijie.*

## ABSTRACT

Micro-Raman spectroscopy and X-ray diffraction have been used to explore the lattice dynamics of Nb-substituted SrBi$_2$(Ta$_{1-x}$Nb$_x$)$_2$O$_9$ (SBTN) crystalline thin films annealed at low temperature, 700ºC. It turned out that SrBi$_2$(Ta$_{1-x}$Nb$_x$)$_2$O$_9$ films consist of fine-grained spherical structures for x=0.1-0.4, while the formation of rod-like grains occurs for x=0.5 due to the stress-induced transformation of the thin film perovskite structure. Moreover, it was revealed that during Nb cationic substitution Aurivillius phase formation was enhanced and become dominated in SBTN thin films and fluorite/pyrochlore phase formation was highly suppressed. We assume that these changes are conditioned by the ferrodistortion occurring in ferroic perovskites, namely by the tilting distortion of (Ta,Nb)O$_6$ octahedra for x=0.2-0.5. The octahedral tilting distortion can change the coordination environment of the A-cite cation, as well as it lowers the SBTN symmetry due to the differences in ionic radius and mass between Ta and Nb in the B-sites, that can lead to significant changes of the SBTN crystal structure. The same nonmonotonic trends were observed for the ferroelectric perovskite phase fraction and remanent polarization on Nb content in SBTN films.

The substituting Nb atoms with a concentration of 10-20 % made it possible to increase the remanent polarization in 3 times and raise the perovskite phase fraction from 66% to 87%. Therefore, obtained results can be used for the production of lead-free thin films with a high remanent polarization under low annealing temperature, being promising for advanced nonvolatile random access ferroelectric memory (NvFeRAM) applications.

# 1. INTRODUCTION

## 1.1. State of the art

Ferroelectric thin films of Aurivillius compounds, such as layered ferroelectric perovskites strontium bismuth tantalate $SrBi_2Ta_2O_9$ (SBT) and niobate $SrBi_2Nb_2O_9$ (SBN) have a relatively high dielectric permittivity, piezoelectric and pyroelectric coefficients, optimal electrooptical properties, intriguing electronic and electrophysical properties, and reveal a robust polarization switching (Shimakawa et al. 1999, Murugan, and Varma 2002, Moure and Pardo 2005). Due to the ability to maintain ferroelectricity in the form of thin films negligible fatigue, low leakage currents, and perovskite Aurivillius compounds are regarded highly-promising candidates for advanced non-volatile ferroelectric random access memories (NvFeRAM) (Jones Jr et al. 1996, . Scott and Paz de Araujo 1989, Paz de Araujo et al. 1995, Noguchi et al. 2003, Yan et al. 1999, Sidsky et al. 2014, Wu et al. 2001).

Recently SBT ferroelectric thin films on Pt electrodes have been extensively investigated for NvFeRAM applications due to their excellent ferroelectric properties. It is well known that the required high post-deposition annealing temperature and low remanent polarization are the two major barriers, which hinder the advanced applications of SBT films. In order to eliminate these two problems, several scientific groups have used various methods, such as forming the solid solutions of $SrBi_2(Ta_xNb_{1-x})_2O_9$ (SBTN) (Chen et al. 1997, Desu et al. 1997, Atsuki et al. 1995), and controlling the oxygen pressure during the annealing process (Atsuki et al. 1995, Ito et al. 1996). Also, it needs to take into account that, in the case of undoped SBT, it is still a challenge to obtain a high remanent polarization at low post-deposition annealing temperatures (Hu et al. 1999, Desu and Vijay 1995). Literature analysis have shown that Nb doping of SBT can reduce the annealing temperature, improve the ferroelectric properties and achieve higher values of remanent polarization and Curie temperature compared to SBT (Chen et al. 1997, Desu et al. 1997, Atsuki et al. 1995, Ito et al. 1996, Hu et al. 1999, Desu and Vijay 1995). Replacing A or B sites in the layered perovskite structure allows effectively improves the dielectric and ferroelectric properties of SBT films. The partial substitution of Ta with Nb in SBTN systems makes it possible to reduce the annealing temperature, improve the ferroelectric properties, achieve higher remanent polarization and Curie temperature compared to undoped SBT (Miura and Tanaka 1998). Since working performances and operating conditions of ferroelectric capacitors for NvFeRAM have a strong relation with the ferroelectric film thickness and post-deposition annealing temperature, one requires to use ferroelectric films not more than (100 - 200) nm thick with relatively high remanent polarization (at least more than 5-10 $\mu C/m^2$) and low post-deposition annealing temperature (e.g., below 750ºC).

In order to reduce the post-deposition annealing temperature, improve the ferroelectric properties and increase the remanent polarization, this work studies $SrBi_2(Ta_{1-x}Nb_x)_2O_9$ thin films with different content of Ta ions substituted by Nb ions (x=0.1-0.5), prepared by sol-gel method. We explore the influence of Ta substitution by Nb on the microstructure and micro-Raman spectrum of sol-gel SBTN thin films annealed at 700°C.

To the best of our knowledge that there are only few works (Osada et al. 2001), where the Raman scattering technique was used for the investigation of SBTN thin films. Our observations clearly imply that the Raman spectroscopy can be used as a rapid and convenient method for characterization of the layered perovskite SBTN structure, and as a sensitive technique for probing lattice vibrational modes, which can provide unique information for identifying the changes in lattice vibrations, as well as the information about the lattice positions occupied by substitutive Nb ions. In addition, Landau-Devonshire approach is used to explain the experimentally observed nonmonotonic variation of perovskite phase fraction and remanent polarization allowing for the possible appearance of finite size effects related with the nanogranular structure of the polycrystalline SBTN films.

### 1.2. SBTN atomic structure and ferroelectric properties

The general formula for the Bi containing layer-type compounds is $Bi_2A_{n-1}B_nO_{3n+3}$, where A denotes the 12-fold coordinated cation in the perovskite sublattice, B represents the octahedral site, bismuth atoms form the rock-salt type interlayer $(Bi_2O_2)^{+2}$ between the perovskite blocks $(A_{n-1}B_nO_{3n+1})$, and "n" is the number of octahedral layers within the perovskite sublattice of the structure. For the stoichiometric $SrBi_2Ta_2O_9$ (n=2) compound there is one complete perovskite sublattice created by the Ta–O octahedra, in which a 12-fold A cation (in our case $Sr^{2+}$) may reside. The $SrBi_2Ta_2O_9$ and $SrBi_2Nb_2O_9$ belong to the $ABi_2B_2O_9$ perovskite-like compounds and have a face-centered orthorhombic unit cell, consisting of $TaO_6$ or $NbO_6$ octahedrons, respectively (see **Figure 1**). These materials have a relatively high Curie temperature of the ferroelectric phase transition ($T_C$), around 335°C (for SBT) (Ching-Prado et al. 1999) and around 440°C (for SBN) (Volanti et al 2007). In the ordered perovskite phase at temperatures below $T_C$ the spontaneous polarization vector lies in the plane of these layers.

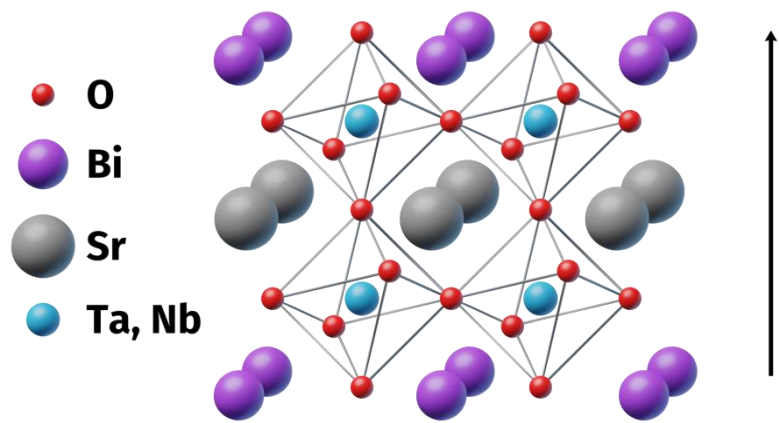

**Figure 1.** A simplified fragment of SBTN structure in the centrosymmetric parent phase above $T_C$. The black arrow shows the possible direction of spontaneous polarization in the ordered perovskite phase at temperatures below $T_C$.

### 1.3. Correlation between SBTN symmetry and active Raman modes

As a first approximation, it is helpful to assign a tetragonal symmetry to the $SrBi_2Ta_2O_9$ and $SrBi_2Nb_2O_9$ above the Curie temperature. It corresponds to a $I_{4/mmm}$ space group. For this symmetry the following Raman and IR modes should be active (Graves et al. 1995):

$$4A_{1g}(R) + 2B_{1g}(R) + 6E_g(R) + 7A_{2u}(IR) + B_{2u}(IR) + 8E_u(IR).$$

At room temperature SBT symmetry is orthorhombic, the space group is $A2_1am$ ($C_{2v}^{12}$), with only slight distortion from tetragonal symmetry. The tetragonal structure is centrosymmetric, so that there is mutual exclusion between infrared and Raman activity. The infrared modes are also Raman active for the orthorhombic structure.

Below the Curie temperature ($T < T_C$), an orthorhombic symmetry is expected in SBT, so that each $E_g$ mode splits into $B_{2g} + B_{3g}$ modes. The splitting at room temperature in these two components must be consistent with the orthorhombic distortion (Kojima et al. 1994, Perez et al. 2003).

SBT has an orthorhombic crystal structure with the $A2_1am$ space group in the ferroelectric phase, which becomes tetragonal in the paraelectric phase with the $I4/mmm$ space group for the temperatures T above 335°C. The inclusion of any cations substituting Sr, Bi or Ta leads to the change in the SBT crystal structure and symmetry depending on the valence and ionic radii of the cations. Nb cation has a smaller ionic radius than Ta, and its location at the Ta-site changes the ionic bond length inside the layered SBT unit cell. Given the mass of Nb and Ta, it is expected that a smaller mass of Nb can shift the vibrational modes to higher frequencies. Because the mass of Nb ion is less than Ta ion, the effective mass of the SBT unit cell becomes smaller after the Ta substitution with Nb, and the effective mass of SBTN cell decreases proportionally to the amount of Nb ions.

## 2. EXPERIMENTAL DETAILS

### 2.1. Materials and methods

The starting product for sol-gel preparation and subsequent formation of the $SrBi_2(Ta_xNb_{1-x})_2O_9$ layer was 0/01 mol/l solution of inorganic metal salts (tantalum, niobium and rare earth chlorides, bismuth and strontium nitrates in toluene). Deposition of SBTN films on $Pt/TiO_2/SiO_2/Si$ substrate was performed by spin-coating of stable solution at the substrate rotation frequency from 800 to 1500 rot/min during 2 to 5 seconds. The adhesion of Pt to conventional dielectrics, such as $SiO_2$, is poor and $TiO_2$ provides an excellent adhesion layer between Pt and $SiO_2$, preventing the formation of platinum silicide. We used the dimethylformamide as a solvent and the monoethanolamine as catalyst. The solvent was removed by multi-stage drying under temperature increase from 80 to 350°C for 6 min. The required thickness of SBT and SBTN films (~100-120 nm) was obtained by layer-on-layer deposition of 2-3 sol-gel layers with heat treatment of each layer at a temperature 700°C. Perovskite SBTN structure was formed during a crystallization post-deposition annealing at 700°C for one hour.

The crystallographic orientation of the films was analyzed by X-ray diffraction (XRD) in the glancing-angle detector scan mode. Raman spectra have been measured using a InVia micro-Raman spectrometer (Renishaw) equipped with a confocal DM2500 Leica optical microscope, a thermoelectrically cooled CCD as a detector, and a laser operating at a wavelength λ = 633 nm excitation from the He–Ne laser. The power of laser used for excitation of Raman spectra is 1.7mW.

Polar and dielectric properties were determined from volt-charge hysteresis loops of Pt/SBTN/Pt/TiO2/SiO2/Si structures with different Nb contents, which were obtained by Sawyer-Tower method using a two-beam oscilloscope. The measurements were carried out in room conditions at the temperature 22°C, relative humidity 70% and atmospheric pressure.

The surface morphology of thin sol-gel films was studied by high-resolution atomic force microscopy (AFM) on SOLVER 47-PRO microscope. The limit of the permissible relative measurement error of linear dimensions in the XY plane is no more than ±1%, the dimensions along the Z axis are not more than ±5%, and the limit of the permissible absolute error of the section of geometric dimensions in the comparator mode (with nominal dimensions of more than 10 nm) is ± 1nm + 0.001L.

Grain sizes and morphology were refined by scanning electron microscopy (SEM) on a Mira high-resolution scanning electron microscope (Tescan, Czech Republic). The error in determining the geometric dimensions of grains on the Mira electron microscope is no more than ±5%, the relative error (on the INCA 350 X-ray energy dispersive spectrometer, OxfordInstrumcnts, UK) in the quantitative analysis mode does not exceed ±5%. Excitation area 0.5 - 1 μm, accelerating voltage 15-20 kV). The number of points for measuring the content of chemical elements for each sample was at least 20.

### 3. EXPERIMENTAL RESULTS AND DISCUSSION

#### 3.1. X-ray and microstructure of SBTN films

Crystallite orientation and phase formation in the SBT and SBTN films were determined by XRD. From the structural point of view, SBT belongs to the bismuth layered perovskite family with the lattice parameters $a$ = 5.531 Å, $b$ = 5.534 Å, $c$ = 25.984 Å in the orthorhombic structure (Rae et al. 1992). The highly anisotropic structure of SBT ($c >> a, b$) results to the strong dependence of its ferroelectric properties on the crystallographic orientation of the film. It has been reported that the ferroelectric properties of SBT thin films in the $a – b$ plane are much stronger than those along the $c$ axis (Wendari et al. 2021, Aleksandrov and Bartolomé 2001, Lonh et al. 2013). This is because the continuous perovskite structure exists in the $a – b$ plane only. Therefore, preparation of SBT thin films with (200) preferential orientation is highly desirable.

**Figure 2** shows the XRD patterns of SBT and SBTN thin films substituted with Nb content x = 0.1, 0.2, 0.3, 0.4 and 0.5. As it can be seen from the figure, SBT film annealed at 700 °C is a mixture of the layered orthorhombic perovskite and fluorite phases. Note, that the phase mixture is typical for low annealing temperatures (below 800 – 900 C), and special measures should be taken to minimize (ideally to exclude) the fluorite phases. In the considered case an optimal Nb doping, such as (10 – 20)%, can be the required measure. It is known that the fluorite phase initially forms at lower temperatures and then transforms either into a bismuth-layered structure, or into a pyrochlore phase SrBi$_x$Ta$_2$O$_9$ ($x<1.2$) after the thermal treatment at 750°C.

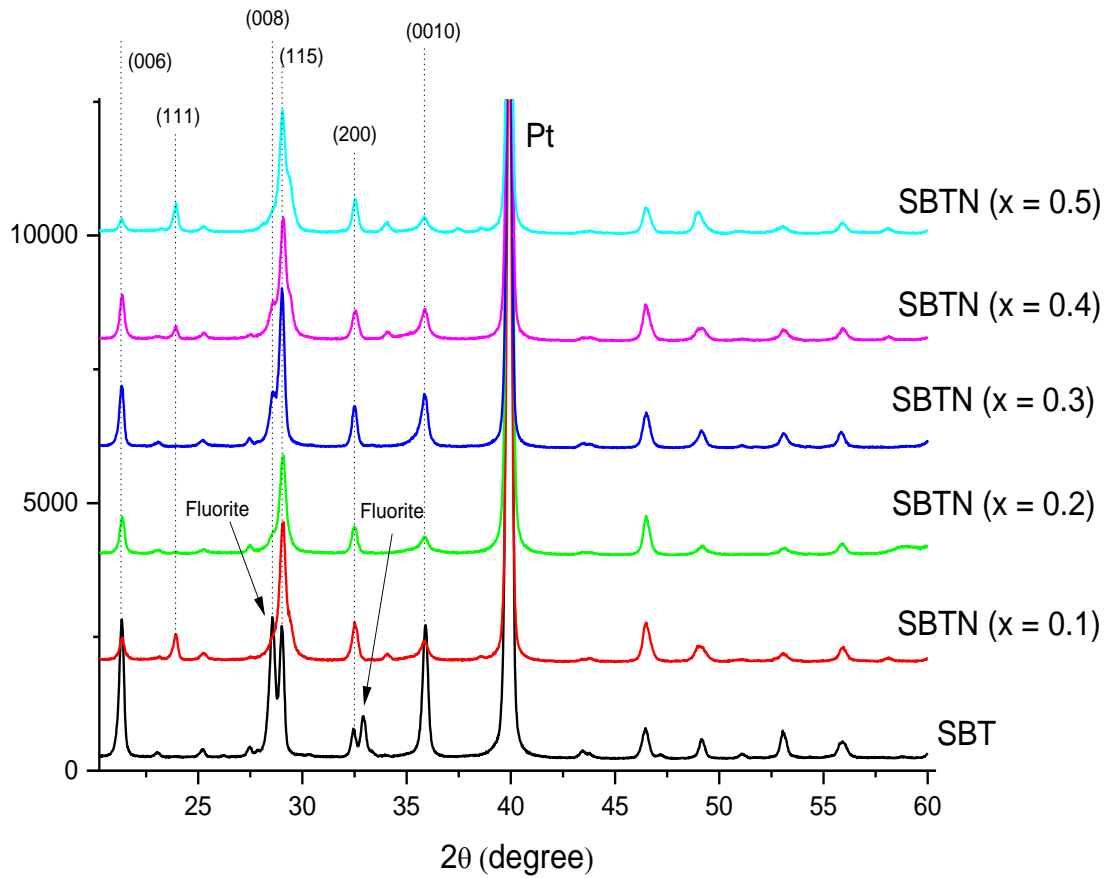

**Figure 2.** XRD patterns of SBT and SrBi$_2$(Ta$_{1-x}$Nb$_x$)$_2$O$_9$ films with various Nb contents.

The intensity of (115) peak increases nonmonotonically with the increase of Nb content x in the SBTN film, and also reveals the nonmonotonic grain growth. In addition, the intensity of (200) peak also increases with the increase of Nb concentration in the film. These results suggest that the structure of SBT thin films can be controlled by the Nb-substitution process. The SBTN films are polycrystalline with a dominant orientation of crystallites along (115) direction; they show the presence of (200) peak responsible for ferroelectric properties. In addition, the peak at the angle $2\theta = 29.4°$ corresponds to a secondary phase and appears at the higher $2\theta$ component of the (115) peak. This peak has been attributed to the distortion of the perovskite unit cell (Desu and Vijay 1995), while authors (Tay et al. 2002, Li et al. 2003) identified it as the non-ferroelectric pyrochlore phase; and the formation of this peak suppresses growth of the ferroelectric (115) peak, which in turn affects the remanent polarization of the SBTN film. This secondary phase is a Bi-deficient pyrochlore phase, which is formed as a result of Bi diffusion into the Pt layer and/extracting of Bi by Pt in the Pt layer. In our case we also observed the decreasing of the ferroelectric

(115) peak for SrBi$_2$(Ta$_{1-x}$Nb$_x$)$_2$O$_9$ thin films with x = 0.5. Also, the broadening of (115) peak profile is observed for SrBi$_2$(Ta$_{1-x}$Nb$_x$)$_2$O$_9$ thin films for x>0.2 due to the increase of the peak intensity at the angle 2θ = 28.5° for x=0.3 and x=0.4, and due to the increase of the peak at the angle 2θ = 29.4° for x=0.4 and x=0.5. Taking into account that the peaks (115) (2θ = 29°) and (200) (2θ = 32,4°) correspond to the ferroelectric phase of SBT (Dinu et al. 1999), it can be assumed that the maximal fraction of the perovskite phase in SrBi$_2$(Ta$_{1-x}$Nb$_x$)$_2$O$_9$ films corresponds to x=0.1-0.2 (see **Figures 2**).

From the AFM pattern of the images in **Figure 3,** it can be seen that in the case of a pure SBT sample, without niobium impurities, the grains stuck together into rather large conglomerates. When niobium was added in a percentage ratio of 10 to 20%, the grain size slightly increased (from 83 to 86 nm). At 30%, a significant decrease in grain size was observed, but at 40%, the grains increased again to 100 nm. We assign these spherical grains to the ferroelectric (115) and (200) phase. The formation of rod-like grains can be observed in SrBi$_2$(Ta$_{1-x}$Nb$_x$)$_2$O$_9$ film when x = 0.5 (see **Figure 3f**) due to the stress-induced structural transformation in thin films, revealed earlier in perovskite thin films (Cao et al. 2017). A broader analysis of the data from Figure 3 can be found in (Sidsky et al. 2017)

Unlikely the rod-shaped grains contribute to the ferroelectric (115) peak, since, as it can be seen from **Figures 2**, its XRD intensity decreases when x = 0.5. We suggest that the rod-like grains correspond to the bismuth-deficient pyrochlore phase with the XRD maximum at the angle 2θ = 29.4°, which intensity increases for the SrBi$_2$(Ta$_{1-x}$Nb$_x$)$_2$O$_9$ with x = 0.5

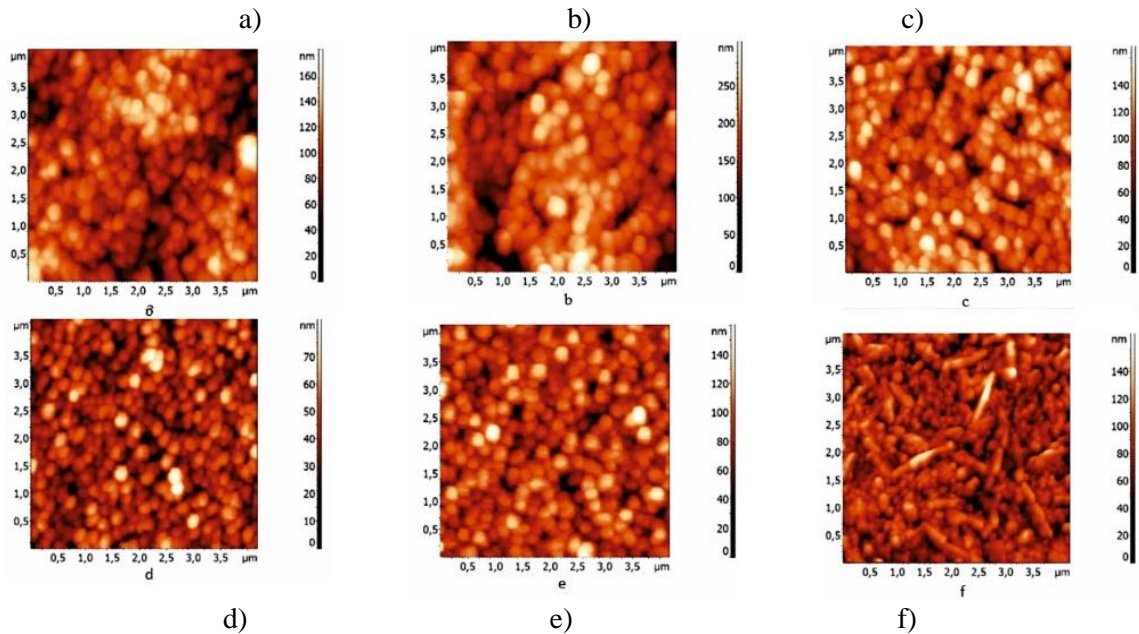

**Figure 3.** AFM images of the SBT and SBTN films surface: **(a)** – SBT film; SrBi$_2$(Ta$_{1-x}$Nb$_x$)$_2$O$_9$ films with x = 0.1 **(b)**, 0.2 **(c)**, 0.3 **(d)**, 0.4 **(e)**, and 0.5 **(f)**.

### 3.2. Raman spectroscopy of SrBi$_2$(Ta$_{1-x}$Nb$_x$)$_2$O$_9$ thin films

Below we analyze Raman spectra of SrBi$_2$(Ta$_{1-x}$Nb$_x$)$_2$O$_9$ thin films with different percent of Ta ions substituted by Nb ions (from 10 to 50 wt.%), annealed at 700 $^0$C. The obtained Raman spectra indicate that the studied films are polycrystalline (see **Figure 4**). Other studies reported that the Raman spectrum of a SBT ceramics exhibits intense bands with maxima at approximately 163, 210, 600, and 805 cm$^{-1}$ which correspond to strong features; and other less intense bands at 319, 356, and 455 cm$^{-1}$ corresponded to weak features (Ortega et al. 2006, Moret et al. 1998). Our results are in overall agreement with these studies, but we register the pronounced Raman bands at approximately 161, 204, 316, 590, and 810 cm$^{-1}$ in the SBT and SBTN films (see **Figure 4**).

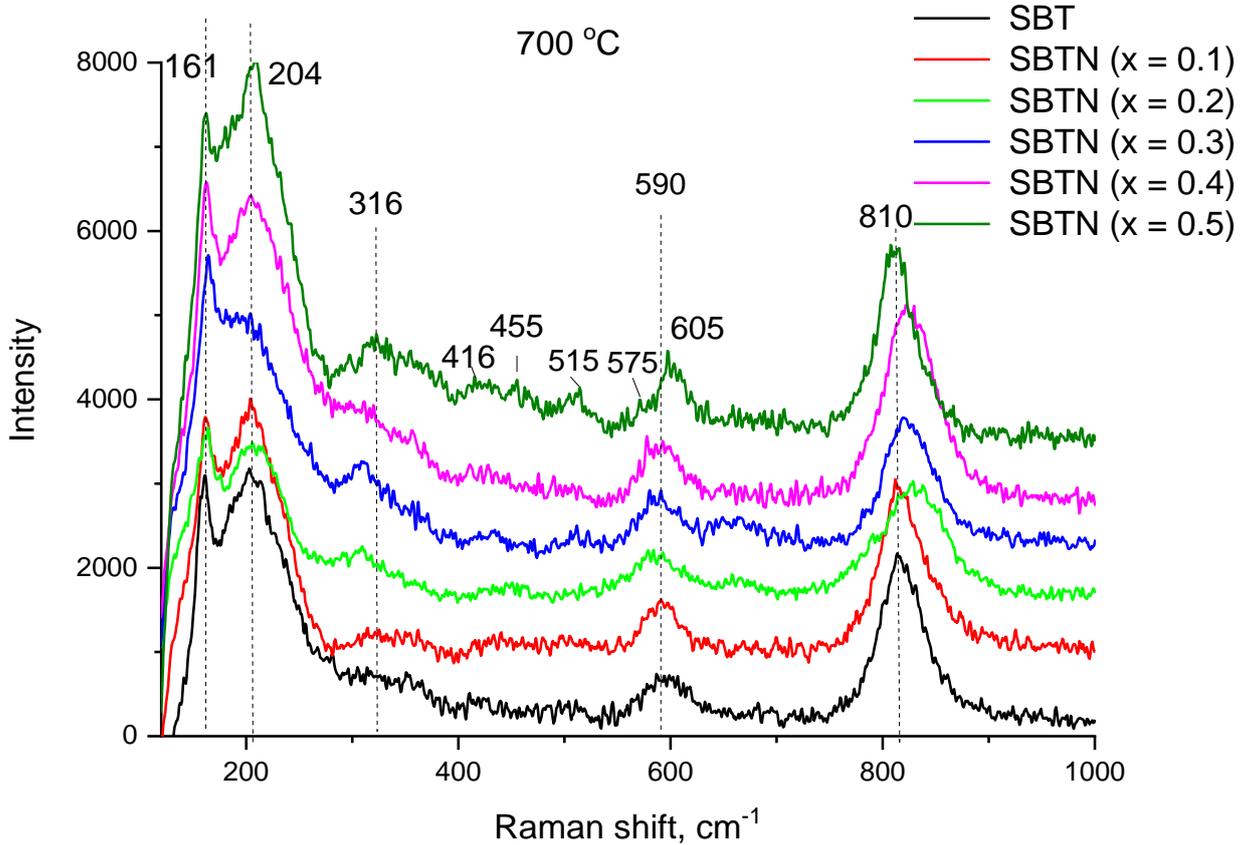

**Figure 4.** Raman spectra of SrBi$_2$(Ta$_{1-x}$Nb$_{1-x}$)$_2$O$_9$ films annealed at 700°C and doped with different wt. % of Nb (x = 0, 0.1, 0.2, 0.3, 0.4 and 0.5).

The band at 161 cm$^{-1}$ is associated with the lattice vibrations of the Ta$^{5+}$ ions at the B-site, namely it corresponds to the TO mode A$_{1g}$ (Liu et al. 2007). The shift of the A$_{1g}$ mode to higher or lower wavenumbers is related with the grain growth. The nondegenerate A$_{1g}$ mode vibrates in the plane perpendicular to the *c*-axis, and its shift to higher frequency occurs because the Ta$^{5+}$ ions are replaced by the lower mass Nb$^{5+}$ ions (Ta$^{5+}$: ≈ 181 amu, 0.68 Å; Nb$^{5+}$: ≈ 93 amu, 0.66 Å).

Relatively small shifts of this mode are observed in SrBi$_2$(Ta$_{1-x}$Nb$_x$)$_2$O$_9$ films, only some slight non-monotonic shifts of this band to higher frequency were observed for x = 0.1 – 0.5 of Nb, as shown in **Figure 5**. The nonmonotonic variation of the frequency shift of the band at 161 cm$^{-1}$ can be related with the transition from the symmetric (Ta-O-Ta) to asymmetric (Ta-O-Nb) bonding with different binding strength under the partial substitution of $Ta^{5+}$ ions with $Nb^{5+}$ ions with x = 0.1 – 0.3; or to the symmetric (Nb-O-Nb) bonding in the oxygen octahedra under $Nb^{5+}$ ions substitution with x = 0.4 – 0.5.

The TO mode at 204 cm$^{-1}$ is related to the vibration of the A-site ion, in our case $Sr^{2+}$. The position of this mode does not change significantly with the addition of Nb (see **Figure 4**). This indicates that Nb does not incorporate into A-site and has only an indirect effect on the site by changing the configuration of the bonds of oxygens during replacing $Ta^{5+}$ by $Nb^{5+}$ cations at the B-site. Note, that the intensities ratio, $I_{161}/I_{204}$, of the bands at 161 cm$^{-1}$ and 204 cm$^{-1}$ reveals a non-monotonic behavior with the increase of Nb content x. The minimum of this ratio is observed for $SrBi_2(Ta_{1-x}Nb_x)_2O_9$ films substituted by Nb (x = 0.5) due to the strong increase of the intensity of the band near 204 cm$^{-1}$, which could be assigned to SrO-group vibration.

The band observed at about 76 cm$^{-1}$ refers to the $E_g$ mode, where the Bi-O atoms in $Bi_2O_2$ layers vibrate out of phase (Kojima 1998). Since the band with maximum at 76 cm$^{-1}$ does not reveal any frequency shifts (see **Figure 5**), except for a slight broadening appearing with the increase of Nb content, we can conclude that $Nb^{5+}$ does not replace $Bi^{+3}$ in $Bi_2O_2$ layers and has only an indirect effect consisting in the changing of the bismuth-oxide layers configuration.

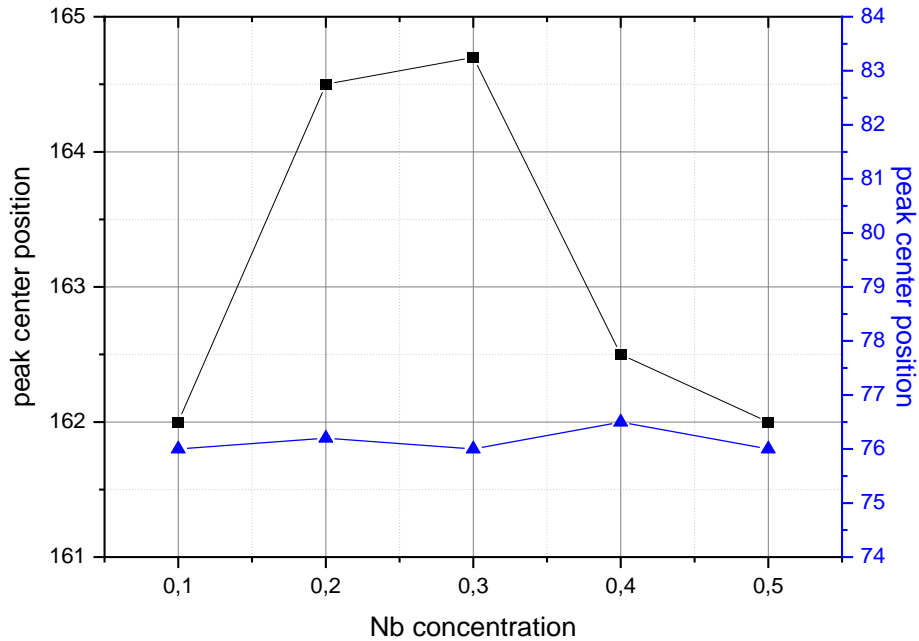

**Figure 5.** Compositional dependence of the band with center position about 162 cm$^{-1}$ (left scale, black squires) and the band with center position about 76 cm$^{-1}$ (right scale, blue triangles) for $SrBi_2(Ta_{1-x}Nb_{1-x})_2O_9$ films with different Nb content (x = 0, 0.1, 0.2, 0.3, 0.4 and 0.5).

The bands in the range (200-300) cm$^{-1}$ can be associated with the O-Ta-O bending and tilting modes (Miura 2003), which are likely to be of $B_{2g}+B_{3g}$ origin due to the degeneracy lifting of the $E_g$ mode. The band around 600 cm$^{-1}$ can be related with the rigid sublattice mode associated with the Ta-O2 or Nb-O2 motions. These modes share their oxygen atom with another Ta or Nb atom, respectively. Modes above 300 cm$^{-1}$ and 590 cm$^{-1}$ bands are the rigid sublattice modes, in which all shifts of anions (O) and cations (Sr, Bi, Ta or Nb) are equal and opposite (Perez et al. 2003, Rae et al. 1992, Tay et al. 2002).

The main significant changes in the Raman spectra of SrBi$_2$(Ta$_{1-x}$Nb$_{1-x}$)$_2$O$_9$ films with different x (from 0 to 0.5) are observed for the band with the maximum around 820 cm$^{-1}$, which is sensitive to the SBTN crystallization degree and corresponds to the stretching mode of (Ta,Nb)O$_6$ octahedra, as it can be observed from **Figure 4**. In the case of pure SBT, the octahedral mode (also known as the breathing mode arising due to volume vibrations) has a frequency of ~810 cm$^{-1}$; it expands and shifts to the higher wavelengths when Ta ions replaced by Nb ions (Chen et al. 1997). A similar blue shift was observed for all Nb concentrations, except x = 0.5 (see **Figure 4**).

In the SBTN films the Raman bands consists of the stretching mode vibrations in SBT ($\approx$ 813 cm$^{-1}$) and SBN ($\approx$ 838 cm$^{-1}$) thin films (Perez et al. 2003), which are shifted from 810 to 838 cm$^{-1}$ under the Nb substitution of Ta. These frequency variations seem to be associated not only with Nb substitution of Ta in the central B ion in the perovskite BO$_6$ octahedra, but also with the number of defects (atoms in non-equilibrium positions, impurities, incomplete solid-state reaction, etc.) existing in the film (Noguchi et al. 2000, Noguchi et al. 2002) and ceramics (Wendari et al. 2019, Wendari et al 2022). The bands at approximately 600 cm$^{-1}$ and 810 cm$^{-1}$ can be attributed to the internal vibration of the (Ta,Nb)O$_6$ octahedron. However, the oxygen ions contributing to the two bands are different. The band at 600 cm$^{-1}$ can be attributed to the vibration of the oxygen ion (O$_2$) at the apex of the TaO$_6$ octahedron. The band at about 810 cm$^{-1}$ can be attributed to the vibration of oxygen ions (O$_4$, O$_5$) in the (Ta,Nb)O$_6$ octahedra (see **Figure 6**). This prominent feature corresponds to the stretching mode (A$_{1g}$ symmetry), and is associated with the symmetric stretching vibrations of the TaO$_6$ and NbO$_6$, octahedrons (Atsuki et al. 1995, Tay et al. 2002). The Raman band around 161cm$^{-1}$ is assigned to the A$_{1g}$ bending mode, involving motions of Bi$^{3+}$ ions perpendicular to the layers.

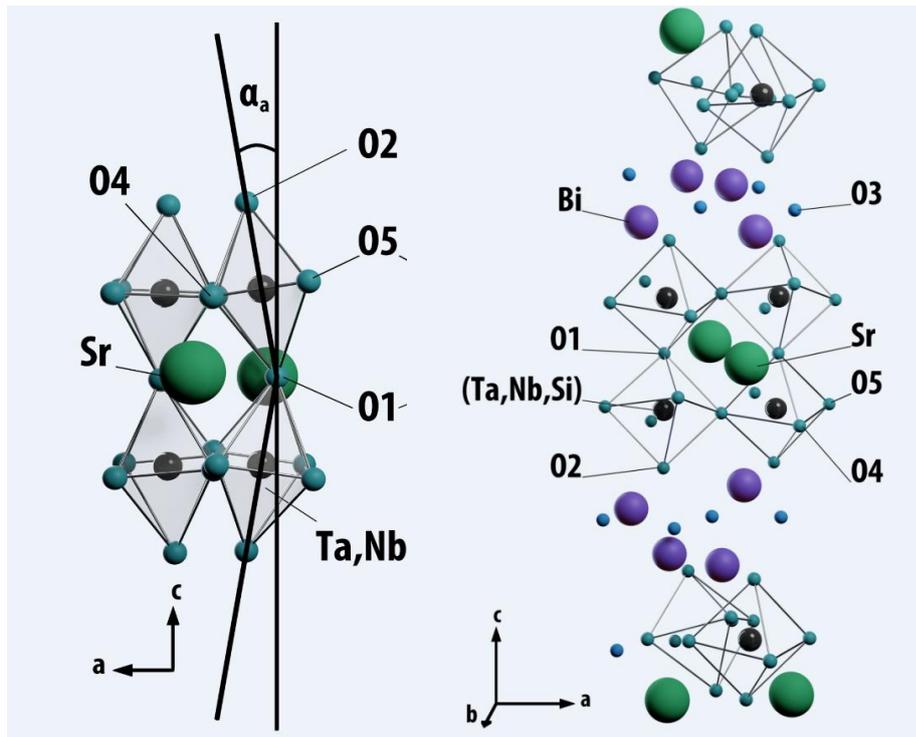

**Figure 6.** Visual structural changes and tilting distortion of (Ta,Nb)O$_6$ octahedra.

For better understanding of this mode nature in SBT and SBTN films, we perform the deconvolution of Raman band with maximum at about 810 cm$^{-1}$ (for SBT) and about 826 cm$^{-1}$ (for SBTN) using the Gaussian fit. Appeared that the band at 810 cm$^{-1}$ for undoped SBT films decomposes into a single maximum, which corresponds to TaO$_6$ mode. An example of deconvolutions of the Raman bands at about 810 cm$^{-1}$ and at about 826 cm$^{-1}$ with and without Nb doping are shown in **Figure S1** in the **Supplement**. The Raman band at about 826 cm$^{-1}$ decomposes into two maxima, at ~(809-813) cm$^{-1}$ and at ~(830 - 837) cm$^{-1}$, which correspond to TaO$_6$ and NbO$_6$ modes, respectively.

Assignment of the dependence of the Nb-O-Nb/Ta-O-Ta ratio on Nb concentration obtained from the deconvolution of the Raman band with maxima at ~(810 - 830) cm$^{-1}$ is shown in **Figure 7.** It can be seen from the figure that the dynamics of changing of the Nb-O-Nb/Ta-O-Ta ratio, FWHM, and the peak center position of the fitting band at ~(810 - 830) cm$^{-1}$ are the same. Also, it needs to note that the increase of Nb ions concentration up to x = 0.5 leads to a significant decrease in the intensity and FWHM of the Gaussian maximum at ~830 cm$^{-1}$, which correspond to NbO$_6$ modes.

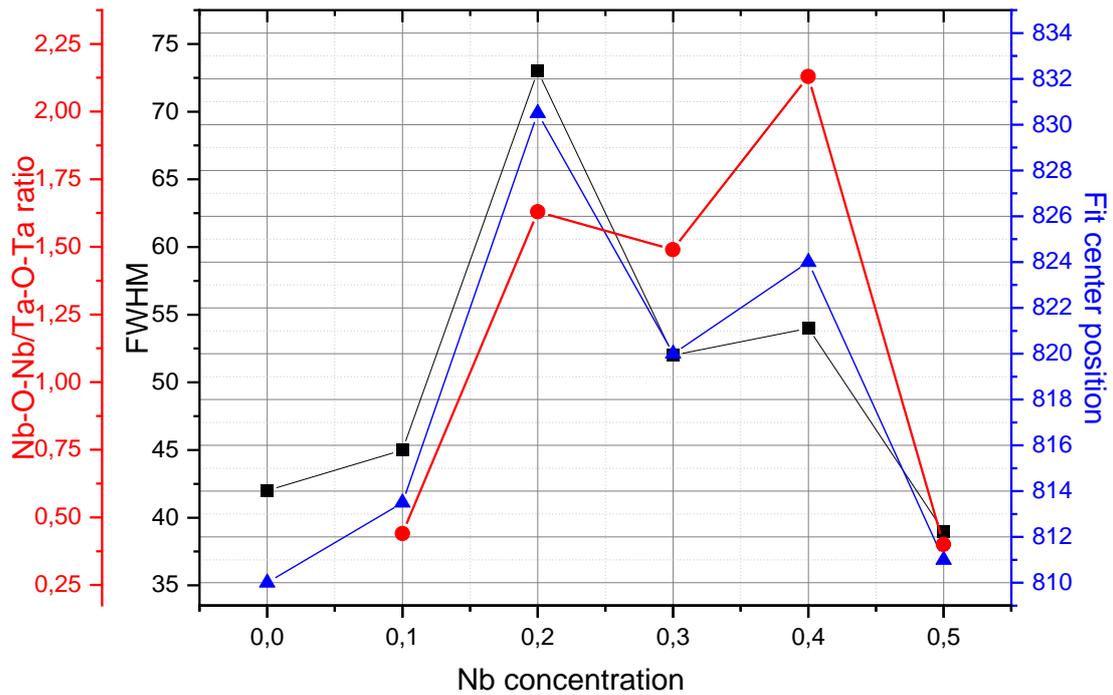

**Figure 7.** Assignment of the dependence of the Nb-O-Nb/Ta-O-Ta ratio obtained from the deconvolution of the Raman band with maxima at about (810-830) cm$^{-1}$ (left scale, red circles), its halfwidth (FWHM) (left scale, black squires), and the position of the fitting center of the band (right scale, blue triangles), depending on the Nb content in SBTN films.

Most important is that we observed a nonmonotonic shift of the Raman band maximum from 810 cm$^{-1}$ (for SBT) to 830 cm$^{-1}$ (for $SrBi_2(Ta_{1-x}Nb_x)_2O_9$, where x = 0.2 – 0.4) and then back to 811 cm$^{-1}$ with increasing Nb content (up to x = 0.5). We assume that these changes are conditioned by the ferro-distortion occurring in perovskites, namely by the tilting distortion of $(Ta,Nb)O_6$ octahedra (**Figure 6**). The octahedra tilting can change the coordination environment of the A-cite cation, as well as it lowers the SBTN symmetry below the cubic symmetry. To resume, the substitution of Nb at Ta-site of SBT has a pronounced influence on the O-Ta-O stretching modes by shifting and splitting the mode frequency at (810 - 830) cm$^{-1}$.

The Raman spectra for x = 0.5 shows a band shift to 605 cm$^{-1}$ and the appearance of the band at 515 cm$^{-1}$. This can be due to the stress-induced transformation in perovskite structure of thin films under high concentration of Nb ions, which effect is similar to the "chemical pressure". The possible transformation is accompanied by the formation of rod-like grains observed in $SrBi_2(Ta_{1-x}Nb_x)_2O_9$ film when x = 0.5 (see **Figure 3f**). The octahedral tilting distortion can change the coordination environment of the A-cite cation, as well as it lowers the SBTN symmetry due to the differences in ionic radius and mass between Ta and Nb in the B-sites, that can lead to significant changes in $SrBi_2(Ta_{1-x}Nb_x)_2O_9$ crystal structure and symmetry when x = 0.5. The same nonmonotonic trends were observed for ferroelectric perovskite phase fraction and remanent polarization on Nb content in SBTN films. They are shown in **Figure 8**. The form of the nonmonotonic changes of the perovskite phase fraction and remanent polarization with increasing Nb content from x = 0 to x = 0.5 strongly resembles the nonmonotonic shift of the Raman band maximum from 810 to 830 cm$^{-1}$ (compare with **Figure 7**). Indeed, the perovskite phase fraction and remanent polarization at first increase with Nb content increase from x = 0 to x = 0.2 followed by a decrease at x = 0.3, increase at x = 0.4, and further decrease for x = 0.4 – 0.5. **Figure 8** also demonstrate the correlation of the perovskite phase with the grain sizes in the sample film. At Nb concentration x=0, the grains stuck together into larger conglomerates, and at x=50, cylindrical structures were formed. The smallest grain size correlates with the smallest perovskite phase at x=30.

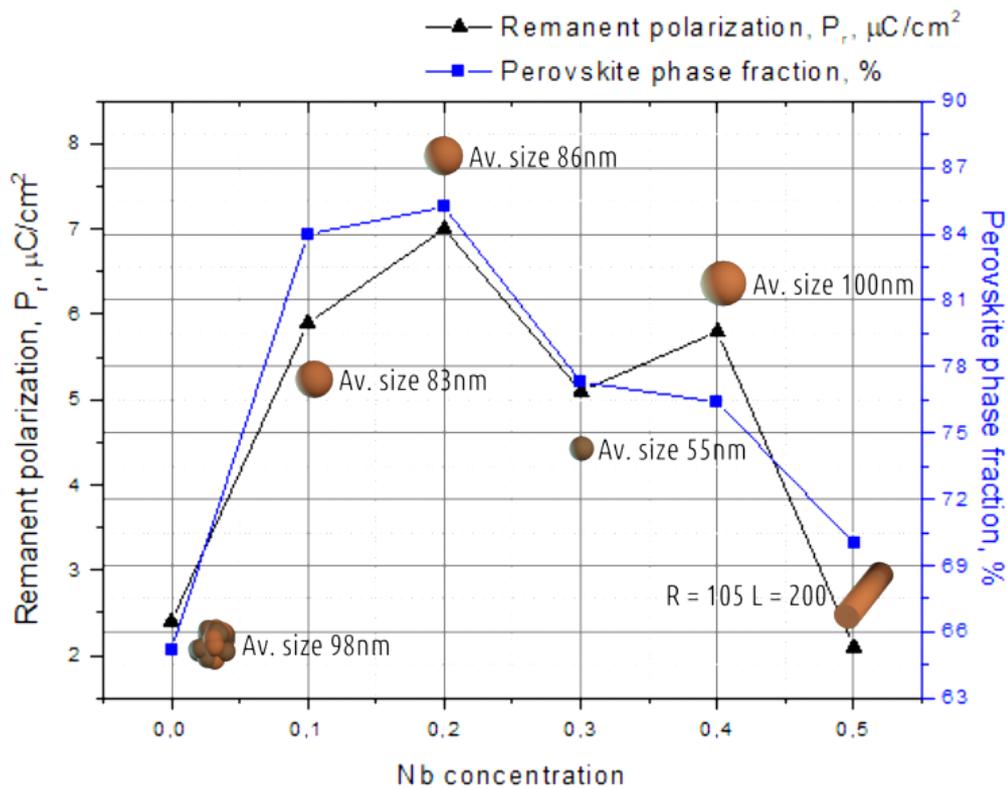

**Figure 8.** Dependence of the remanent polarization (left scale, black squires) and the perovskite phase fraction (right scale, blue triangles) on the Nb content in SBTN thin films. Small images show the grain shape and average size.

Hence, the oxygen octahedra tilting can explain the nonmonotonic changes of the perovskite phase fraction and remanent polarization with increasing Nb. To describe quantitatively the experimentally observed nonmonotonic dependence of the ferroelectric perovskite phase fraction and remanent polarization on Nb content in SBTN films, we use Landau-Devonshire approach in the next subsection.

## 3.3. Landau-Devonshire approach for the description of the perovskite phase fraction and polar properties of SBTN

The finite size effect on phase transitions in SrBi$_2$Ta$_2$O$_9$ nanoparticles have been investigated by *in situ* Raman scattering by Yu et al. (2003), and by thermal analysis and Raman spectroscopy by Ke et al (2007). The finite size and Nb substitution effects impact the phase diagrams, polar and dielectric properties of the SBTN nanogranular films (Morozovsky et al. 2015, Eliseev et al. 2016). These effects can be described using the phenomenological Landau-Devonshire approach, taking into account the effect of Nb substitution on the structural ferrodistortion of SBTN and the charge state of this (nominally isovalent) dopant.

The addition of Nb increases the Curie temperature $T_C$, and thus improves the ferroelectric properties. It is reasonable to assume that at low concentrations of Nb the dependence $T_C(x)$ decomposes into a series of Nb content $x$, $T_C(x) = T_C\left(1 + \sum_{i=1}^{n} a_i x^i\right)$, where $a_i$ are unknown decomposition coefficients. However, with a further increase in $x$, the tilting and disorder in the system of oxygen octahedrons increases, because Nb and Ta have different force matrices. The degree of disorder is the maximum for x = 50 %, and with further increase, we gradually move to the ferroelectric SBN. The effect of variable valence is also not excluded, although it is unlikely to play a key role. The increase in disorder and tilting affect the magnitude of spontaneous polarization due to nonlinear interaction with the structural subsystem and the correlation of the structural and polar order parameters. We assume that the nonlinearity also decomposes into a series of $x$ (because it is at least an even interaction), $\beta(x) = \beta_t\left(1 + \sum_{i=1}^{n} b_i x^i\right)$, where $b_i$ are unknown decomposition coefficients. In this simplest model we have the following $x$-dependent coefficients in the SBTN Landau-Devonshire free energy:

$$F = \frac{\alpha^*(T,x)}{2} P^2 + \frac{\beta(x)}{4} P^4 + \frac{\gamma}{6} P^6 - PE \qquad (1)$$

Here **P** is the polarization, **E** is the electric field, and α, β, γ are the coefficients of expansion into *P*-series. Due to the finite size effect related with the films nanogranular structure, the "effective" coefficient $\alpha(T, x)$ can depend on the average sizes of the grains, which are modeled by nanoellipsoids with semi-axes $R$ and $L$ (Eliseev et al. 2016):

$$\alpha^*(T,x) = \alpha(T,x) + \frac{n_d}{\varepsilon_0[\varepsilon_b n_d + \varepsilon_e(1-n_d) + n_d(D/\lambda)]}, \quad \alpha(T,x) = \alpha_T[T - T_C(x)]. \qquad (2a)$$

Here $\varepsilon_b$ and $\varepsilon_e$ are the dielectric permittivity of ferroelectric background (Tagantsev and Gerra 2006) and external media respectively, $n_d = \frac{1-\xi^2}{\xi^3}\left(\ln\sqrt{\frac{1+\xi}{1-\xi}} - \xi\right)$ is the depolarization factor, $\xi = \sqrt{1 - (R/L)^2}$ is the eccentricity ratio of ellipsoid with a shorter semi-axes $R$ and longer semi-axis $L$ (Landau and Lifshic 1984); and $D$ is the ellipsoid semi-axis ($R$ or $L$) in the direction of spontaneous polarization $P$.

The dielectric stiffness $\alpha(T, x)$ and the nonlinearity $\beta(x)$ can be written as follows:

$$\alpha(T,x) = \alpha_t\left(T - T_C\left(1 + \sum_{i=1}^{n} a_i x^i\right)\right), \quad \beta(x) = \beta_t\left(1 + \sum_{i=1}^{n} b_i x^i\right). \qquad (2b)$$

Where $T$ is the absolute temperature, $T_C$ is the Curie temperature, the coefficients $\alpha_t$ and $\beta_t$ are positive. Coefficient $\gamma \geq 0$, and the γ-term can be neglected for the case of the second order phase transitions, and then the spontaneous polarization and coercive field are equal to:

$$P_S(T,x) = \sqrt{-\frac{\alpha(T,x)}{\beta(x)}}, \qquad E_C(T,x) = -\frac{2}{3}\alpha(T,x)\sqrt{\frac{-\alpha(T,x)}{3\beta(x)}}. \qquad (3a)$$

Knowing $P_S(T,x)$ and $E_C(T,x)$ from the experiment, we can determine the coefficients of Landau-Devonshire expansion $\alpha(T,x)$ and $\beta(x)$:

$$\alpha(T,x) = -\frac{3\sqrt{3}}{2}\frac{E_C(T,x)}{P_S(T,x)}, \qquad \beta(x) = -\frac{\alpha(T,x)}{P_S^2(T,x)} \equiv \frac{3\sqrt{3}}{2}\frac{E_C(T,x)}{P_S^3(T,x)}. \qquad (3b)$$

Since $T_C \approx (314 - 426)$ °C for SBT and $T_C \approx (414 - 494)$ °C for SBN (Cho et al. 2004), the Curie-Weiss constant $C_{CW}$ can be determined from the value of the linear dielectric constant of SBT at room temperature, for which the formula $\frac{1}{2\varepsilon_0 \alpha_t (T_C - T)}$ and its value (180-200) for a bulk homogeneous SBT and SBN, respectively. We obtain the value $C_{CW} = (2 - 0,55) \cdot 10^5$ K. The dependences of the Landau expansion coefficients $\alpha(T,x)$ and $\beta(x)$ on the Nb concentration can be adjusted by means of polynomial functions by fitting the experimental values of spontaneous polarization and coercive field.

The experimental data (Sidsky et al. 2017) from **Table 1** were used to fit the experimental results. The results of fitting $P_S(T,x)$ and $E_C(T,x)$ using polynomial functions of the 4-th order and Eqs.(3a) are shown in **Figures 9a** and **9b**, respectively. **Figures 9c** and **9d** show the dependences of the Landau free energy coefficients $\alpha(T,x)$ and $\beta(x)$ on the Nb concentration (as a percentage), calculated from equations (3b) for SBTN at room temperature. As follows from **Figure 9a**, SBTN films with Nb content of x=0.1 and 0.2 have the highest spontaneous polarization (Sidsky et al. 2017).

**Table 1**. Dependence of the main ferroelectric characteristics of SBT and SBTN on Nb content.

| Parameters | SBT | x value in SrBi$_2$(Ta$_{1-x}$Nb$_x$)$_2$O$_9$ | | | | |
|---|---|---|---|---|---|---|
| Content of Nb, x | 0 | 0.1** | 0.2 | 0.3 | 0.4 | 0.5 |
| Perovskite phase, % | 65 | 86 | 85 | 77 | 76 | 70 |
| Spontaneous polarization, μC/cm$^2$  * | 2.68 | 7.90 | 7.00 | 5.10 | 5,80 | 2,10 |
| Coercive field, kV/cm | 50 | 90 | 70 | 65 | 80 | 100 |
| Bulk dielectric susceptibility at 293 K, dimensionless | 180 - 200 | 180 - 200 | 180 - 200 | 180 - 200 | 180 - 200 | 180 - 200 |

* Obtained values of spontaneous polarization are relatively small compared to some other Aurivillius compounds, due to the phase structure of studied thin film. Indeed, the nonpolar pyrochlore phase is present

in films, which fraction decreases the spontaneous polarization. The highest polarization (7.9 μC/cm² at x=0.1) corresponds to the highest fraction of perovskite phase.

** Data for polarization and coercive field of SBT-0.1 are taken from Ref. (Sidsky et al. 2017). Other data are collected from our measurements.

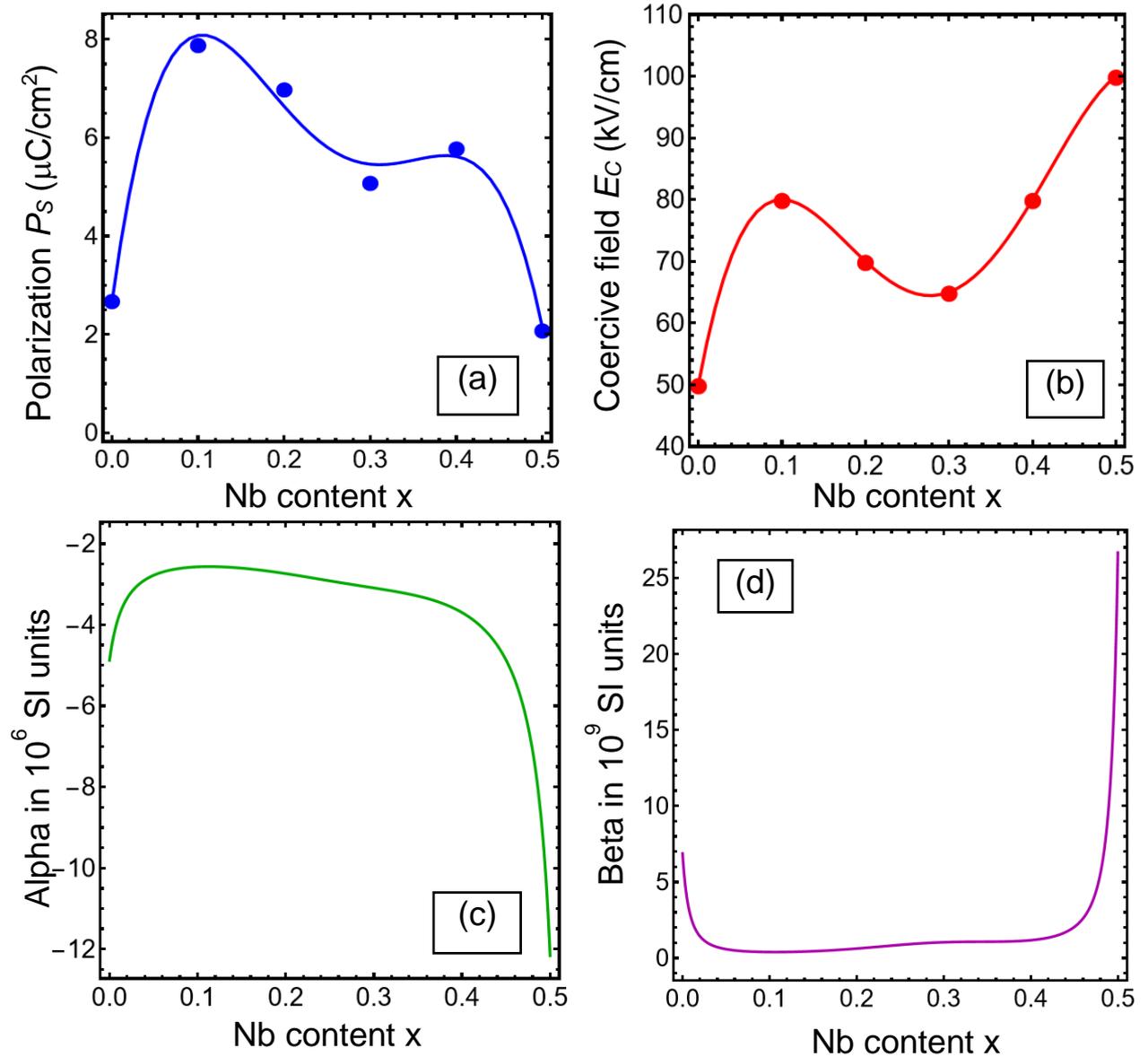

**Figure 9.** Dependences of the spontaneous polarization $P_S$ **(a)**, coercive field $E_c$ **(b)**, Landau expansion coefficients $\alpha(T,x)$ **(c)** and $\beta(x)$ **(d)** on Nb content $x$, calculated from Eqs.(3) for SBTN films at different Nb content $x$ at room temperature. The blue and red circles in graphs (a) and (b), are the experimental results for the $P_S$ and $E_c$, respectively, listed in **Table 1**.

As follows from **Figure 10a**, SBTN film with x=0.1 and 0.2 is as close as possible to the perovskite structure and the surface of the film has a finer grain structure (average grain size about 83-86 nm). An increase above x=0.2 in the Nb content leads to a decrease in the fraction of the perovskite phase, as is evidenced by the expansion of the XRD line (115) and a decrease in its intensity at the angle 2θ ~ 28.9 deg shown in **Figures 2**. The change in the fraction of the perovskite phase with the introduction of the Nb ions into the SBT matrix can be explained by the change in the parameters of the crystal lattice, its bond strength and rigidity, and the surface energy of the material. The appearance of the crystal structure anisotropy leads to a decrease in the perovskite phase fraction in the SBTN film, with an increase above x=0.2 in the Nb content. An increase to x=0.5 of Nb in the SBTN film leads to the formation of quasi-cylindrical grains with an average size of $R = (105 \pm 3)$ nm, $L = (200 \pm 3)$ nm, which leads to the deterioration of the ferroelectric properties ($Ps$ =2.1 μC/cm$^2$, see **Figure 9a**). The average grain size increases with the increase of Nb content above x=0.3 and it is about (100-120) nm (the crystallite size in this case is about 23 nm). The SEM of SBTN films, shown in **Figure 10b,** and corresponding AFM images, shown in **Figure 3**, demonstrate the formation of elongated quasi-ellipsoidal grains for x=0.5 of Nb.

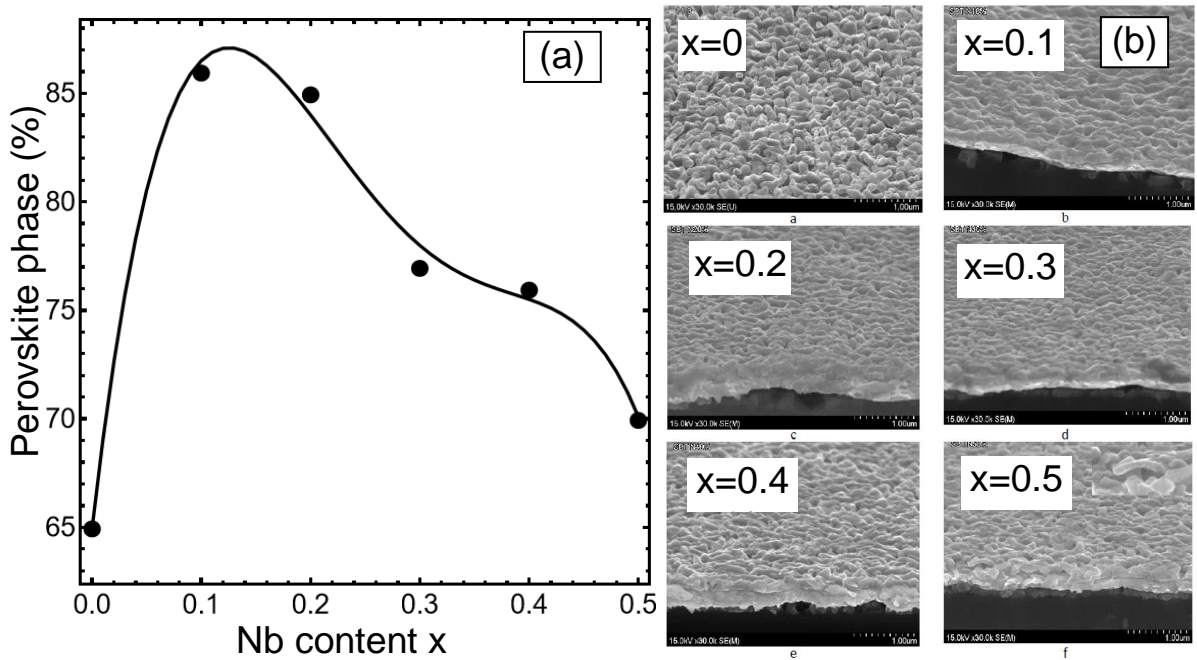

**Figure 10**. **(a)** The dependence of the SBTN perovskite phase fraction (in %) on the Nb content x, calculated from equations (3a) for SBTN. Black circles in the part (a) and SEM images in the part (b) are taken from experimental results (Sidsky et al. 2017). **(b)** SEM images of SBTN films for different content x of Nb, which are written in legends, are experimental results (Sidsky et al. 2017).

It is reasonable to assume that $P_S(T,x)$ is proportional to the fraction of the perovskite phase, and this is confirmed by comparing the shape of solid curves in **Figures 9a** and **10a**. This comparison is an additional basis for the applicability of Eqs.(3) to the analysis of experimental results in order to determine the previously unknown dependences of the Landau free energy coefficients $\alpha(T,x)$ and $\beta(x)$ on the Nb content $x$. Hence, the comparison of theoretical results with experimental ones, shown in **Figures 9** and **10**, reveals a linear correlation between the spontaneous polarization of nanogranular SBTN films and the perovskite phase fraction in them. This is an additional basis for the applicability of the Landau-Devonshire approach to the analysis of experimental results for SBTN in order to determine the previously unknown dependences of the Landau expansion coefficients on the concentration of substitutes and dopants. To resume, the proposed Landau-Devonshire approach describes the experimentally observed nonmonotonic dependence of the ferroelectric perovskite phase fraction and remanent polarization on Nb concentration. Important that the form of the nonmonotonic dependences shown in **Figures 9a** and **10a** strongly resembles the dependence of the shift of the Raman band maximum from 810 cm-1 to 830 cm-1 shown in Figure 7.

Our results demonstrate that the nonmonotonic Raman band shift, which strongly correlates with the ferroelectric perovskite phase fraction and remanent polarization value, is conditioned by the ferrodistortion occurring in perovskite structure, namely by the transient tilting distortion of (Ta,Nb)O6 octahedra at x=0.2-0.5. In principle, this result can be an indicator on the previously unknown structural phase transition induced by Nb-doping, which is likely of the ferrodistortive-type. Hence, our finding can explain the nonmonotonic concentration dependence in many other ferroics, such as e.g., HfxZr1-xO2 thin films, which are dielectric - ferroelectric - antiferroelectric when x varies from 1 to 0 (see e.g. (T. Mikolajick et al. 2021)).

## 4. CONCLUSION

In summary, polycrystalline SBT and $SrBi_2(Ta_{1-x}Nb_x)_2O_9$ thin films with x = 0.1, 0.2, 0.3, 0.4 and 0.5 were prepared by sol-gel method and annealed at relatively low temperature, 700°C, and characterized by XRD. The relative intensity of the (115), (006) and (200) XRD peaks, which correspond to the crystal structure and ferroelectric phase, nonmonotonically increases with the increase of Nb content.

Analysis of SEM and AFM images allows to conclude that $SrBi_2(Ta_{1-x}Nb_x)_2O_9$ films consist of fine-grained spherical structures, if the percent of Ta ions substituted by Nb ions is x = 0.1 – 0.4, while the formation of rod-like grains occurs when the percent of Ta ions substituted by Nb ions reaches x = 0.5 due to the stress-induced transformation of perovskite structure in the thin film.

Micro-Raman spectroscopy was used to study the lattice vibrational modes, and to explore the internal structure, polycrystalline nature, and crystallization behavior of the SBTN thin films with Nb ions at Ta-sites. Analysis of the Raman spectra leads to the following conclusions:

a). The vibration mode at 810 cm$^{-1}$ observed in the case of the undoped SBT sample, is also known as the octahedral stretching mode (or breathing mode). The mode shows a significant shift to higher frequencies (up to 830 cm$^{-1}$) and broadening upon substitution of $Ta^{5+}$ ions with $Nb^{5+}$ ions up to x = 0.4. The effect is explained by the presence of vibrations of both types, Ta-O-Ta and Nb-O-Nb, in the structure. The substitution of Nb at Ta-site shows a significant splitting of O-Ta-O octahedral stretching mode. Unexpectedly, we observed a nonmonotonic shift of the Raman band maximum from 810 cm$^{-1}$ (for pure SBT) to around 830 cm$^{-1}$ (for $SrBi_2(Ta_{1-x}Nb_x)_2O_9$ with x = 0.2 – 0.4) and then back to 811 cm$^{-1}$ with increasing Nb content from up to x = 0.5. We assume that these changes are conditioned by the ferrodistortion occurring in ferroic perovskites, namely by the tilting distortion of (Ta,Nb)$O_6$ octahedra happened at x = 0.2 – 0.5. The octahedral tilting distortion can change the coordination environment of the A-cite cation, as well as it lowers the SrBi$_2$(Ta$_{1-x}$Nb$_x$)$_2$O$_9$ symmetry due to the differences in ionic radius and mass between Ta and Nb in the B-sites, that can lead to significant changes of the SrBi$_2$(Ta$_{1-x}$Nb$_x$)$_2$O$_9$ crystal structure and symmetry when x = 0.5. The revealed octahedral tilting distortion can explain the nonmonotonic changes of the perovskite phase fraction and remanent polarization with increasing Nb content from x = 0 to x = 0.5, in particular their initial increase with Nb content increase up to x = 0.2 followed by a decrease at x = 0.3, then increase at x = 0.4 and further decrease for x = 0.4 – 0.5.

b). The nonmonotonic variation of the frequency shift of the Raman band at 161 cm$^{-1}$ is related with the transition from symmetric (Ta-O-Ta) to the asymmetric (Ta-O-Nb) bonding with different binding strength under the partial substitution of Ta$^{5+}$ ions with Nb$^{5+}$ ions with x = 0.1 and 0.3, or symmetric (Nb-O-Nb) bonding in the oxygen octahedrons under Nb$^{5+}$ ions substitution with x =0.4 and more. This behavior is explained in terms of smaller ionic radii of Nb than Ta at B-site.

c). The Raman band at about 76 cm$^{-1}$ refers to the E$_g$ mode, where the Bi-O groups in Bi$_2$O$_2$ layers in the same plane vibrate out of phase. The band maximum shows the almost absence of any frequency shift with the increasing of Nb content. Hence, we concluded that Nb$^{5+}$ does not replace Bi$^{+3}$ in Bi$_2$O$_2$ layers and has only indirect influence consisting in changing the configuration of the bismuth-oxide layers. Since Nb is pentavalent and has been embedded in an octahedral lattice, the low-frequency modes do not undergo significant changes.

d). It has been shown experimentally, that the changes of the perovskite phase fraction and remanent polarization values, which appear with Nb content increase from x = 0 to x = 0.5 in SrBi$_2$(Ta$_{1-x}$Nb$_x$)$_2$O$_9$ thin films, are the same as the nonmonotonic dependences of both factors, the Raman band maximum shift from 810 cm$^{-1}$ to (833-837) cm$^{-1}$, and its halfwidth, on the concentration of Nb.

Performed theoretical analysis shows the correlations between the structural phase changes and Raman bands shift related with symmetric stretching modes of the (Ta,Nb)O$_6$ octahedra. In particular, Landau-Devonshire approach describes the experimentally observed nonmonotonic dependence of the ferroelectric perovskite phase fraction and remanent polarization on Nb content. The form of the nonmonotonic dependence strongly resembles the nonmonotonic shift of Raman band maximum from 810 cm$^{-1}$ to 830 cm$^{-1}$ appearing with Nb content increase.

Thus, the combination of the micro-Raman spectroscopy with X-ray diffraction, ferroelectric measurements and Landau-Devonshire theoretical approach allows to establish the role of Nb doping and to reveal the physical origin of strong correlations between the lattice dynamics, microstructure, phase composition and ferroelectric properties of the SBTN thin films annealed at relatively low temperature.

We have shown that the substitution of Ta atoms by Nb atoms with a concentration of 10-20 % made it possible to increase the remanent polarization in 3 times and raise the perovskite phase fraction from 66% to 87%. Thus, the obtained results can be useful for the preparation of thin films with a high remanent polarization under low annealing temperatures, which are promising for the advanced applications in NvFeRAM.

**Data availability statement:** The datasets used and analyzed during the current study are available from the corresponding author on reasonable request. For convenience, we added a supplementary docx file with all original raw data for **Figures 2, 4, 5, 7, 8**, which can be opened in Origin software by a mouse double-click. Theoretical results, presented in **Figures 9** and **10a,** are made using Mathematica 12.2, corresponding notebook, SBT-SBTN.nb, with all codes is supplemented and/or available from the corresponding author on reasonable request.

**Acknowledgements.** A.N.M. expresses deepest gratitude to Dr. N.V. Morozovsky (NASU) for useful discussions and valuable suggestions. The work (O.M.F., A.D.Y., T.V.T., O.P.B. and A.N.M.) is supported by the National Academy of Sciences of Ukraine and the European Union's Horizon 2020 research and innovation programme under the Marie Skłodowska-Curie grant agreement No 778070, Transferr and by the grant of the Ministry of Education and Science of Ukraine "Influence of dimensional effects on the electrophysical properties of graphene-ferroelectric nanostructures". This work (L.W., A.V.S., V.V.S) is supported in part by the Natural Science Foundation of China under Grant 62222108.

**Authors' contribution.** O.M.F., A.D.Y., T.V.T., O.P.B. conducted experiments and wrote the experimental part of the work. L.W., A.V.S and V.V.S. prepared samples. A.N.M. performed calculations and wrote the theoretical part of the work. All authors worked on the results discussion and manuscript improvement.

**Conflicts of Interest Statement.** The authors have no competing interests to declare that are relevant to the content of this article. All authors certify that they have no affiliations with or involvement in any organization or entity with any financial interest or non-financial interest in the subject matter or materials discussed in this manuscript.

# References

Aleksandrov, K. S., & Bartolomé, J. (2001). Structural distortions in families of perovskite-like crystals. Phase Transitions: A Multinational Journal, 74(3), 255-335.

Atsuki, T., Soyama, N., Yonezawa, T., & Ogi, K. O. K. (1995). Preparation of Bi-based ferroelectric thin films by sol-gel method. Japanese journal of applied physics, 34(9S), 5096.

Cao, S. G., Li, Y., Wu, H. H., Wang, J., Huang, B., & Zhang, T. Y. (2017). Stress-induced cubic-to-hexagonal phase transformation in perovskite nanothin films. Nano letters, 17(8), 5148-5155.

Chen, T. C., Thio, C. L., & Desu, S. B. (1997). Impedance spectroscopy of $SrBi_2Ta_2O_9$ and $SrBi_2Nb_2O_9$ ceramics correlation with fatigue behavior. Journal of materials research, 12(10), 2628-2637.

Ching-Prado, E., Pérez, W., Reynés-Figueroa, A., Katiyar, R. S., Ravichandran, D., & Bhalla, A. S. (1999). Raman study of $SrBi_2Ta_2O_9$ thin films. Ferroelectrics Letters Section, 25(3-4), 97-102.

Cho, J. A., Park, S. E., Song, T. K., Kim, M. H., Lee, H. S., & Kim, S. S. (2004). Dielectric and piezoelectric properties of nonstoichiometric $SrBi_2Ta_2O_9$ and $SrBi_2Nb_2O_9$ ceramics. Journal of electroceramics, 13(1), 515-518.

Desu, S. B., Joshi, P. C., Zhang, X., & Ryu, S. O. (1997). Thin films of layered-structure $(1-x)SrBi_2Ta_2O_9 - xBi_3Ti(Ta_{1-y}Nb_y)O_9$ solid solution for ferroelectric random access memory devices. (1997) Applied physics letters, 71(8), 1041-1043.

Desu, S. B., & Vijay, D. P. (1995). Novel fatigue-free layered structure ferroelectric thin films. Materials Science and Engineering: B, 32(1-2), 75-81.

Dinu, R., Dinescu, M., Pedarnig, J. D., Gunasekaran, R. A., Bäuerle, D., Bauer-Gogonea, S., & Bauer, S. (1999). Film structure and ferroelectric properties of in situ grown $SrBi_2Ta_2O_9$ films. Applied Physics A, 69(1), 55-61.

Eliseev, E. A., Semchenko, A. V., Fomichov, Y. M., Glinchuk, M. D., Sidsky, V. V., Kolos, V. V., ... & Morozovska, A. N. (2016). Surface and finite size effects impact on the phase diagrams, polar, and dielectric properties of (Sr, Bi) $Ta_2O_9$ ferroelectric nanoparticles. Journal of Applied Physics, 119(20), 204104.

Graves, P. R., Hua, G., Myhra, S., & Thompson, J. G. (1995). The Raman modes of the Aurivillius phases: temperature and polarization dependence. Journal of Solid State Chemistry, 114(1), 112-122.

Hu, G. D., Wilson, I. H., Xu, J. B., Cheung, W. Y., Wong, S. P., & Wong, H. K. (1999). Structure control and characterization of $SrBi_2Ta_2O_9$ thin films by a modified annealing method. Applied physics letters, 74(9), 1221-1223.

Ito, Y., Ushikubo, M., Yokoyama, S., Matsunaga, H., Atsuki, T., Yonezawa, T., & Ogi, K. (1996). New low temperature processing of sol-gel $SrBi_2Ta_2O_9$ thin films. Japanese journal of applied physics, 35(9S), 4925.

Jones Jr, R. E., Zurcher, P., Jiang, B., Witowski, J. Z., Lii, Y. T., Chu, P., ... & Gillespie, S. J. (1996). Electrical characterization of $SrBi_2Ta_2O_9$ thin films for ferroelectric non-volatile memory applications. Integrated Ferroelectrics, 12(1), 23-31.

Ke, H., Jia, D. C., Wang, W., & Zhou, Y. (2007). Ferroelectric phase transition investigated by thermal analysis and Raman scattering in $SrBi_2Ta_2O_9$ nanoparticles. In Solid State Phenomena (Vol. 121, pp. 843-846). Trans Tech Publications Ltd.

Kojima, S. (1998). Optical mode softening of ferroelectric and related bismuth layer-structured oxides. Journal of Physics: Condensed Matter, 10(20), L327.

Kojima, S., Imaizumi, R., Hamazaki, S., & Takashige, M. (1994). Raman scattering study of bismuth layer-structure ferroelectrics. Japanese journal of applied physics, 33(9S), 5559.

Landau, L. D., & Lifshitz, E. M. (1984). 8: Electrodynamics of Continuous Media. Butterworth-Heinemann.

Li, A. D., Wu, D., Ling, H. Q., Yu, T., Liu, Z. G., & Ming, N. B. (2003). Role of interfacial diffusion in $SrBi_2Ta_2O_9$ thin-film capacitors. Microelectronic engineering, 66(1-4), 654-661.

Liu, G. Z., Wang, C., Gu, H. S., & Lu, H. B. (2007). Raman scattering study of La-doped $SrBi_2Nb_2O_9$ ceramics. Journal of Physics D: Applied Physics, 40(24), 7817.

Long, C., Fan, H., & Ren, P. (2013). Structure, Phase Transition Behaviors and Electrical Properties of Nd Substituted Aurivillius Polycrystallines Na0. 5Nd x Bi2. 5–x Nb2O9 (x= 0.1, 0.2, 0.3, and 0.5). Inorganic Chemistry, 52(9), 5045-5054.

Mikolajick, T., Slesazeck, S., Mulaosmanovic, H., Park, M. H., Fichtner, S., Lomenzo, P. D., ... & Schroeder, U. (2021). Next generation ferroelectric materials for semiconductor process integration and their applications. Journal of Applied Physics, 129(10), 100901.

Miura, K. (2003). Electronic and Structural Properties of Ferroelectric $SrBi_2Ta_2O_9$, $SrBi_2Nb_2O_9$ and $PbBi_2Nb_2O_9$. Journal of the Korean Physical Society, 42(9), 1244.

Miura, K. M. K., & Tanaka, M. T. M. (1998). Difference in the electronic structure of $SrBi_2Ta_2O_9$ and $SrBi_2Nb_2O_9$. Japanese journal of applied physics, 37(2R), 606.

Moret, M. P., Zallen, R., Newnham, R. E., Joshi, P. C., & Desu, S. B. (1998). Infrared activity in the Aurivillius layered ferroelectric SrBi$_2$Ta$_2$O$_9$. Physical Review B, 57(10), 5715.

Morozovsky, N. V., Semchenko, A. V., Sidsky, V. V., Kolos, V. V., Turtsevich, A. S., Eliseev, E. A., & Morozovska, A. N. (2015). Effect of annealing on the charge–voltage characteristics of SrBi$_2$(Ta$_x$Nb$_{1-x}$)$_2$O$_9$ films. Physica B: Condensed Matter, 464, 1-8.

Murugan, G. S., & Varma, K. B. R. (2002). Microstructural, dielectric, pyroelectric and ferroelectric studies on partially grain-oriented SrBi$_2$Ta$_2$O$_9$ ceramics. Journal of electroceramics, 8(1), 37-48.

Moure, A., & Pardo, L. (2005). Ferroelectricity in aurivillius-type structure ceramics with n= 2 and (SrBi$_2$Nb$_2$O$_9$)$_{0.35}$(Bi$_3$TiNbO$_9$)$_{0.65}$ composition. Journal of electroceramics, 15(3), 243-250.

Noguchi, Y., Kitamura, A., Woo, L. C., Miyayama, M., Oikawa, K., & Kamiyama, T. (2003). Praseodymium-modified SrBi$_2$Ta$_2$O$_9$ with improved polarization properties at low electric field. Journal of applied physics, 94(10), 6749-6752.

Ortega, N., Bhattacharya, P., & Katiyar, R. S. (2006). Enhanced ferroelectric properties of multilayer SrBi$_2$Ta$_2$O$_9$/SrBi$_2$Nb$_2$O$_9$ thin films for NVRAM applications. Materials Science and Engineering: B, 130(1-3), 36-40.

Noguchi, Y., Miwa, I., Goshima, Y., & Miyayama, M. (2000). Defect control for large remanent polarization in bismuth titanate ferroelectrics–doping effect of higher-valent cations–. Japanese Journal of Applied Physics, 39(12B), L1259.

Noguchi, Y., Miyayama, M., Oikawa, K., Kamiyama, T., Osada, M., & Kakihana, M. (2002). Defect engineering for control of polarization properties in SrBi$_2$Ta$_2$O$_9$. Japanese journal of applied physics, 41(11S), 7062.

Osada, M., Kakihana, M., Mitsuya, M., Watanabe, T., & Funakubo, H. (2001). Raman spectroscopic fingerprint of ferroelectric SrBi$_2$Ta$_2$O$_9$ thin films: A rapid distinction method for fluorite and pyrochlore phases. Japanese Journal of Applied Physics, 40(8B), L891.

Perez, W., Das, R. R., Dobal, P. S., Yuzyuk, Y. I., Bhattacharya, P., & Katiyar, R. S. (2003). Effect of cationic substitution on Raman spectra of SrBi$_2$Ta$_2$O$_9$ ceramics and thin films. MRS Online Proceedings Library (OPL), 784.

Rae, A. D., Thompson, J. G., & Withers, R. L. (1992). Structure refinement of commensurately modulated bismuth strontium tantalate, $Bi_2SrTa_2O_9$. Acta Crystallographica Section B: Structural Science, 48(4), 418-428.

Scott, J. F., & Paz de Araujo, C. A. (1989). Ferroelectric memories. Science, 246(4936), 1400-1405.

De Araujo, C. A. P. A. Z., Cuchiaro, J. D., McMillan, L. D., Scott, M. C., & Scott, J. F. (1995). Fatigue-free ferroelectric capacitors with platinum electrodes. Nature, 374(6523), 627-629.

Shimakawa, Y., Kubo, Y., Nakagawa, Y., Kamiyama, T., Asano, H., & Izumi, F. (1999). Crystal structures and ferroelectric properties of $SrBi_2Ta_2O_9$ and $Sr_{0.8}Bi_{2.2}Ta_2O_9$. Applied Physics Letters, 74(13), 1904-1906.

Sidsky V. V., Semchenko A. V., Kolos V. V., Petlitsky A. N., Solodukha V. A., Kovalchuk N. S. (2017) Influence of processing conditions on the structure and ferroelectric properties of SBTN films obtained by the sol-gel method. Problems of physics, mathematics and technology, no. 1 (30), 17-21.

Sidsky, V. V., Semchenko, A. V., Rybakov, A. G., Kolos, V. V., Turtsevich, A. S., Asadchyi, A. N., & Strek, W. (2014). La3+-doped SrBi2Ta2O9 thin films for FRAM synthesized by sol-gel method. Journal of Rare Earths, 32(3), 277-281.

Tagantsev, A. K., & Gerra, G. (2006). Interface-induced phenomena in polarization response of ferroelectric thin films. Journal of applied physics, 100(5), 051607.

Tay, S. T., Huan, C. H. A., Wee, A. T. S., Liu, R., Goh, W. C., Ong, C. K., & Chen, G. S. (2002). Substrate temperature studies of $SrBi_2(Ta_{1-x}Nb_x)_2O_9$ grown by pulsed laser ablation deposition. Journal of Vacuum Science & Technology A: Vacuum, Surfaces, and Films, 20(1), 125-131.

Volanti, D. P., Cavalcante, L. S., Paris, E. C., Simões, A. Z., Keyson, D., Longo, V. M., ... & Hernandes, A. C. (2007). Photoluminescent behavior of $SrBi_2Nb_2O_9$ powders explained by means of $\beta$-$Bi_2O_3$ phase. Applied physics letters, 90(26), 261913.

Wendari, T. P., Arief, S., Mufti, N., Suendo, V., Prasetyo, A., Baas, J., & Blake, G. R. (2019). Synthesis, structural analysis and dielectric properties of the double-layer Aurivillius compound Pb1-2xBi1. 5+2xLa0. 5Nb2-xMnxO9. Ceramics International, 45(14), 17276-17282.

Wendari, T. P., Arief, S., Mufti, N., Insani, A., Baas, J., & Blake, G. R. (2021). Structure-property relationships in the lanthanide-substituted PbBi2Nb2O9 Aurivillius phase synthesized by the molten salt method. Journal of Alloys and Compounds, 860, 158440.

Wendari, T. P., Arief, S., Mufti, N., Blake, G. R., Baas, J., Suendo, V., ... & Zulhadjri, Z. (2022). Lead-Free Aurivillius Phase Bi2LaNb1. 5Mn0. 5O9: Structure, Ferroelectric, Magnetic, and Magnetodielectric Effects. Inorganic Chemistry, 61(23), 8644-8652.

Wu, Y., Forbess, M. J., Seraji, S., Limmer, S. J., Chou, T. P., & Cao, G. (2001). Impedance study of $SrBi_2Ta_2O_9$ and $SrBi_2(Ta_{0.9}V_{0.1})_2O_9$ ferroelectrics. Materials Science and Engineering: B, 86(1), 70-78.

Yan, Y., Al-Jassim, M. M., Xu, Z., Lu, X., Viehland, D., Payne, M., & Pennycook, S. J. (1999). Structure determination of a planar defect in $SrBi_2Ta_2O_9$. Applied physics letters, 75(13), 1961-1963.

Yu, T., Shen, Z. X., Toh, W. S., Xue, J. M., & Wang, J. (2003). Size effect on the ferroelectric phase transition in $SrBi_2Ta_2O_9$ nanoparticles. Journal of applied physics, 94(1), 618-620.